\documentclass[12pt]{article}
\usepackage{graphicx}
\usepackage{amsfonts}

\title{Opportunities at the Mathematics/Future Cities Interface\footnote{University of Strathclyde
                           Mathematics and Statistics 
                  Research Report~10 (2014).
 A revised version of this article 
 will appear in 
 SIAM News (Society of Industrial and Applied Mathematics).
}
}

\author{
Peter Grindrod\thanks{Department of Mathematical Institute, University of Oxford, UK} \and 
Desmond J. Higham\thanks{Department of Mathematics and Statistics, University of Strathclyde, UK} \and  
Robert S. MacKay\thanks{
Mathematics Institute and Centre for Complexity Science,
University of Warwick,
UK
}}

\begin{document}
\maketitle

\date{}

\begin{abstract}
We make the case for mathematicians  and statisticians to stake their claim in the fast-moving and high-impact research field that is becoming known as Future Cities. After assessing the Future Cities arena, we provide some illustrative challenges where mathematical scientists can make an impact.
\end{abstract}

\begin{figure}
  \begin{center}
        \scalebox{0.3}{\includegraphics{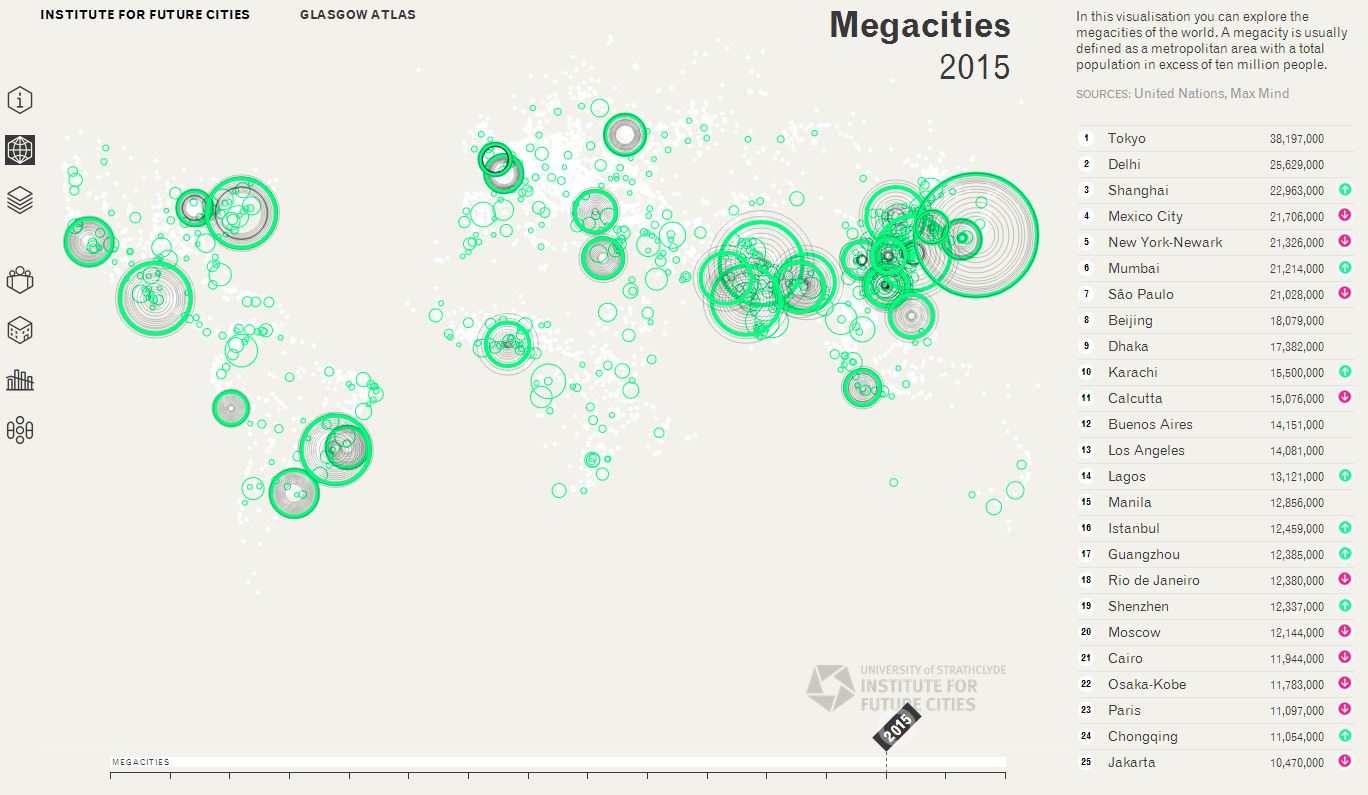}}
          \end{center}
                      \caption{
                 Megacities---population exceeding 10 Million.
                 Credit: c.LUSTlab/Institute for Future Cities, University of Strathclyde.
                 Reproduced with permission.
              }
          \label{Fig.Mega}
        \end{figure}

\begin{figure}
  \begin{center}
        \scalebox{0.3}{\includegraphics{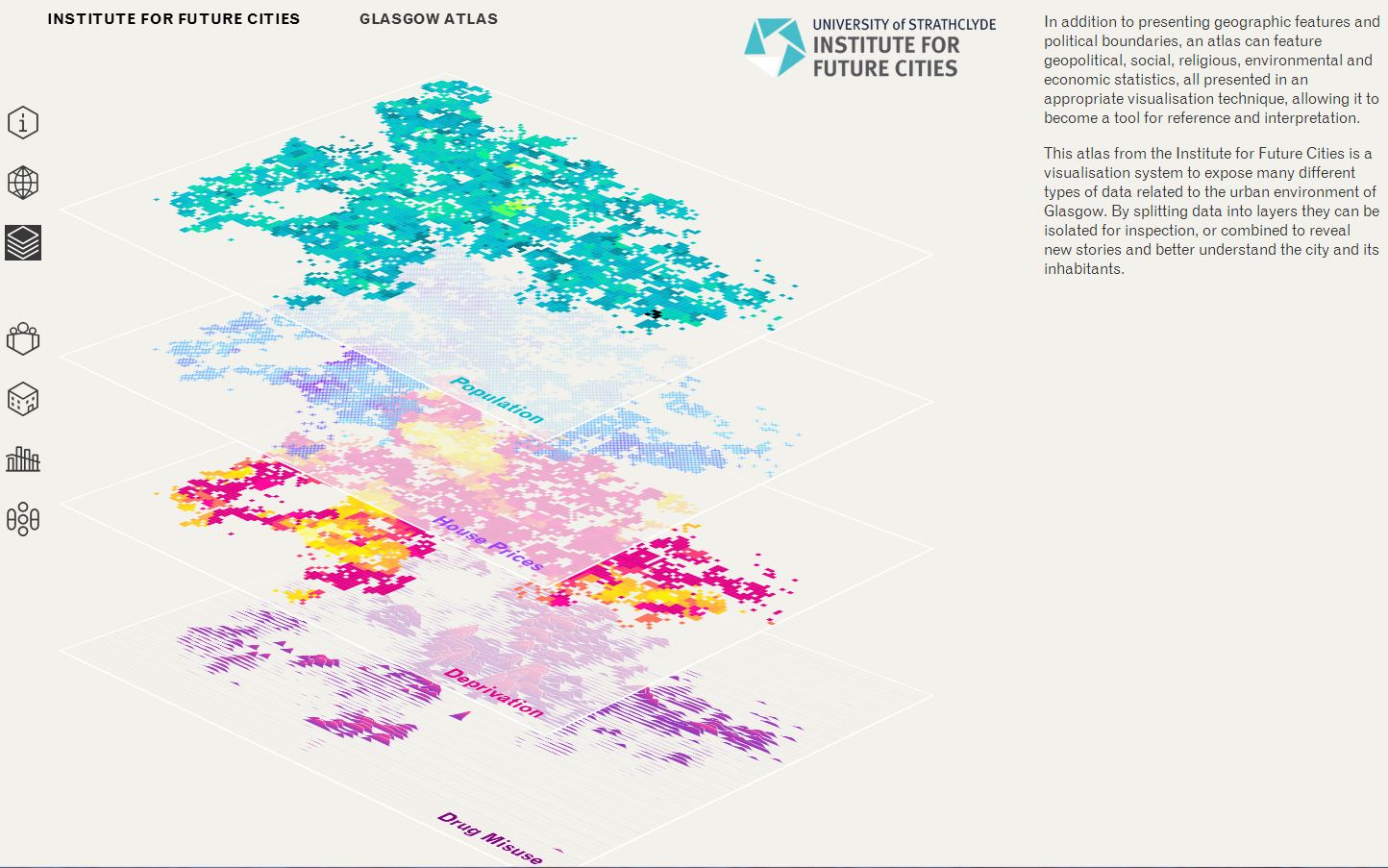}}
          \end{center}
                      \caption{
                 Multiplex visualization of population density, housing costs,         
                   level of deprivation and level of drug misuse across 
                  the city of Glasgow, UK.
                 Credit: c.LUSTlab/Institute for Future Cities, University of Strathclyde.
                 Reproduced with permission.
              }
          \label{Fig.Glasgow}
        \end{figure}

\begin{figure}
  \begin{center}
        \scalebox{0.25}{\includegraphics{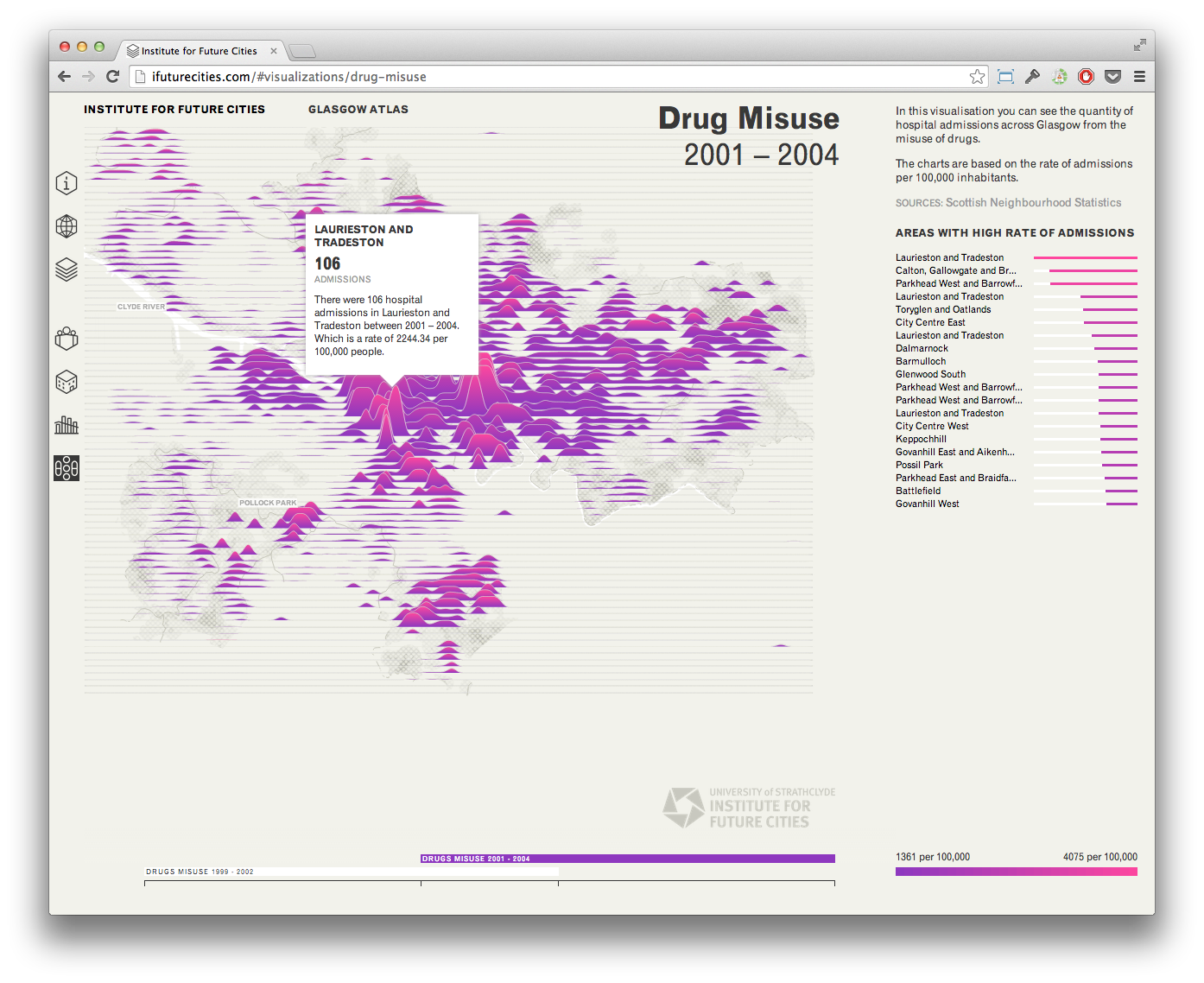}}
          \end{center}
                      \caption{
                 A screen-shot extract from the data streams in Figure~\ref{Fig.Glasgow}, 
                 showing reported level of drug
                 misuse across the city of Glasgow.
                 After discretization, for example based on city regions, 
                 combining the levels in Figure~\ref{Fig.Glasgow}
                 naturally leads to a three-dimensional tensor, where two dimensions
                 represent spatial coordinates and the third dimension indexes the data sources.
                 Time-dependency in the data  would add a fourth dimension.
                 Extracting commonalities and differences, and summarizing 
                 patterns, can be cast in terms of tensor factorization---for example, 
                 generalizing the well known matrix-level Singular Value Decomposition (SVD).
                 Note, however, that those four dimensions are not comparable---any 
                 results should be insensitive to the order in which we label the data streams, 
                 but, for most purposes, we should not reorder points in time or space.
                 Credit: c.LUSTlab/Institute for Future Cities, University of Strathclyde.
                 Reproduced with permission.
              }
          \label{Fig.Drug}
        \end{figure}

More than half of the world's population lives in a city, and this proportion is estimated to 
reach 60\% by 2030 and 70\% by 2050 [World Health Organization, Urban Population Growth, July, 2014]. 
See Figure~\ref{Fig.Mega} for a graphic showing our current 
\lq\lq megacities.\rq\rq 
Thanks to the proliferation of smart devices and interconnected services, cities are gushing with 
data, much of which relates to human behavior. 
City life generates data streams around on-line social media, telecommunication, 
geo-location, crime, health, transport, air quality, energy, utilities, weather, CCTV, wi-fi usage, 
retail footfall and satellite imaging. 
Viewing urban centres as \lq\lq Living Labs\rq\rq\  is a powerful new concept that is inspiring novel research 
leading to improved wellbeing and economic growth. 
We argue here that mathematicians can make an impact at the heart of this emerging 
interdisciplinary field, where hypotheses about human behavior must be quantified and tested against 
vast data sets
and where decisions and interventions should be based on quantitative, testable predictions.
Further, the rapid growth of large-scale, disparate, multi-resolution data sets is driving new research challenges 
for applied and computational mathematicians, drawing on
 hot topic areas such as dynamic and multiplex networks  (see Figures~\ref{Fig.Glasgow} and 
\ref{Fig.Drug}) 
\cite{multirevb,multirev},
multiscale modelling and simulation \cite{Battymulti}, 
uncertainty quantification \cite{SmithUQ} and 
sparse tensors 
\cite{papalexakis2013scoup,ScLoVaKoXX}.

Since the terms are relatively new and open to interpretation, we are using 
\lq\lq Future Cities\rq\rq\ here as a catch-all to cover topics that may equally well be classified under 
\lq\lq Smart Cities\rq\rq\  or  \lq\lq Urban Analytics.\rq\rq\ However, for reference, we note that 
Batty et al. 
\cite[page 481]{Battysmart} define a smart city to be \lq\lq a city in which ICT 
[Information and Communication Technologies] is merged with traditional infrastructures, 
coordinated and integrated using new digital technologies\rq\rq\ 
and the recent report by 
Arup 
\cite{Arupmarket}
makes an attempt to distinguish between smart and future versions: 
\lq\lq There is still confusion in the market as to the distinction between Smart city solutions and 
Future city solutions. 
Future city solutions are innovative physical projects which are often but not exclusively associated with 
low carbon economies. 
Smart city solutions apply digital technologies to address social, environmental and economic goals.\rq\rq\  
The Future City research arena we envisage is inherently interdisciplinary, covering the physical and 
social sciences, engineering,  business, law, and, in particular, dealing with issues of  privacy and ethics. 
At the risk of buzzword overload, we also note that Future Cities is a topic that has strong overlaps with 
other big picture themes, including Data Science, 
Big Data, 
Complexity, 
Planet Earth, Digital Economy, 
Internet of Things and Computational Social Science.

Many urban centers across the world are becoming active in the Future Cities space, 
with governments and funding agencies showing strong support for these developments. 
Focussing on the authors' home institutions, Glasgow City Council beat 
30 other cities to win a \pounds 24M Future Cities Demonstrator competition, 
funded by the UK government's innovation agency, the Technology Strategy Board; 
within this award, the Institute for Future Cities at the University of Strathclyde 
is developing a Digital Observatory that will allow public domain access to data 
generated in Glasgow and elsewhere. 
Future Cities is also one of the four strategic themes for Strathclyde's \pounds 89M 
Technology and Innovation Centre, a hub for academic research and industrial collaboration, 
and the university offers a Master’s degree in Leadership for Global Sustainable Cities. 
The University of Oxford's 
Engineering and Physical Sciences Research Council (EPSRC) Centre for Doctoral Training in new Industrially Focused 
Mathematical Modelling has a strong data/analytics/technology component and its 
Said Business School hosts the Institute for New Economic Thinking. 
The University of Warwick, which has designated Sustainable Cities as one of its 
Global Priority Programmes, houses the Warwick Institute for the Science of Cities,
and offers an EPSRC Centre for Doctoral Training in Urban Science.  
The University of Warwick is also a partner in the Centre for
Urban Science and Progress 
(CUSP), a public-private research collaboration using 
New York City as a laboratory and classroom
[Link to Sidebar 1], 
and in its recently announced branch 'CUSP London'.

\bigskip

\noindent
\textbf{Sidebar 1}\\
\begin{center}
***********************************
\end{center}

Quoting from CUSP’s website at 
\verb5http://cusp.nyu.edu/about/5
\begin{quote}
\lq\lq CUSP will instrument New York City and use existing data from a network of 
agencies to transform the city into a living laboratory and classroom. 
It will make sense of the vast amount of data it collects to help cities around 
the world become more productive, more livable, more equitable, and more resilient.\rq\rq
\end{quote}

\begin{center}
***********************************
\end{center}

\bigskip

Looking further afield, 
Horizon 2020, the biggest ever 
 European Union research and innovation programme, 
chose 
Societal Challenges as one of its three pillars,  in which 
 a 
100 Million Euro call for research projects
is listed 
under the theme
Smart Cities and Communities.
The UK's science and engineering research council, EPSRC,
released a draft 
Strategic Plan in July 2014 that 
listed \lq\lq designing and building future cities\rq\rq\ as one of 
seven key challenges for the global economy, and
the Technology Strategy Board 
chose Future Cities for one of its seven 
\emph{Catapult Centres} 
[Link to Sidebar 2]. 

\bigskip

\noindent
\textbf{Sidebar 2}\\

\begin{center}
***********************************
\end{center}

 A Catapult is a physical centre where businesses, scientists and engineers 
work together to develop ideas into new products and services.
The London-based Future Cities Catapult was established in June 2013; see,
 \verb5https://futurecities.catapult.org.uk/5.
An illustrative  project is \emph{Sensing London}, which focusses on 
\begin{description}
\item[Data Collection:] deploying a range of air pollution sensors across 
        Hyde Park, Brixton, Enfield and Elephant \& Castle.
 \item[Data Mashing:] overlaying and integrating data, and applying state-of-the-art 
 algorithmics, modelling and visualisation to generate new insights. 
  \item[Trialling innovation:] for example, building a virtual  \lq\lq asthma-guard\rq\rq\ 
        to let asthmatics know  
         in real time where it is safe to walk in the city.
\end{description}

\begin{center}
***********************************
\end{center}

\bigskip

The report 
\cite{Arupmarket},
 commissioned by the UK Department for Business, Innovation and Skills, 
considered opportunities for UK industry in smart city technology across five urban market 
verticals---energy, water, transport, waste and assisted living---estimating 
a global market of \$408 Billion by 2020. 
As we prepared this article, Cisco announced plans to open a \$30 Million Global Internet of Everything Centre in Barcelona, focusing on smart cities.

To be more concrete about opportunities for
the mathematical sciences community
we now focus briefly on 
recent developments and prospects in dynamical systems and 
in networks. 
Our discussion is, of course, biased towards our own research interests.

Macro-scale observations have revealed
scaling laws that relate city population size to 
other attributes, such as 
energy consumption, 
household income and
patent production,
and important distinctions 
 have been 
drawn between linear, sublinear and superlinear  
growth
\cite{boundaries2013,scaling2014}.
Explanatory, micro-scale models based on \lq\lq hidden\rq\rq\ 
laws must of course be consistent with such observations.
Long-time dynamics and stability are key issues 
in the modelling of complex urban systems, as are sensitivities to 
parameter choices, including thresholds due to resource limitations 
\cite{citygrowth2011}.
In principle, 
good 
mathematical models can be used to 
map out the ranges of possible behavior, helping us to understand 
whether we might be observing a phenomenon that is constrained within a single 
domain of attraction (whilst there are others as yet unseen) with a very low 
probability of breaking out, or else we might be observing a
trajectory of 
a chaotic process, where the qualitative  macroscopic behaviour is
predictable but the quantitative evolution of specific individuals is not (due to sensitive dependence and
instability driven disruptions).
In modelling terms, 
it may not be appropriate to surgically extract the city from its surroundings, and an open 
model, subject to a range of external influences, may be more realistic.
Phenomena of interest may then be subject to 
persistent cycling or boiling---never approaching quiescence \cite{aperiodic2013}.

Digital interactions in an urban setting can naturally be represented as graphs, or networks,  
but the links between nodes in the system typically have an important  time-dependent feature: 
who just texted whom, who just 
logged in to which free wifi zone,
who just reported a crime at which location?
Two of us have written previously in SIAM News about how a dynamic view of classical
concepts in graph theory has led to useful new algorithms \cite{SIAMNewspeople}.
However, alongside the data-driven issue of extracting and summarizing 
information from network observations, there is an equally compelling challenge to
derive models that describe the underlying dynamics.
Representing a network as a time-dependent matrix, $A(t)$,  whose $(i,j)$ element quantifies the current level
of interaction between nodes $i$ and $j$, we can formalize concepts from the social
sciences to derive suitable laws of motion [Link to Sidebar 3].
In an urban context, where dynamic interactions take place on many levels between 
a range of parties, it is natural to think of dynamic models that operate across many layers, with 
the dynamics on one layer (say,  the evolution of 
attitudes towards healthy lifestyle) 
coupled to the dynamics on another 
(say, the reach of a social media campaign). 
Moreover, with the advent of smartphones and 
GPS, we can now 
monitor geographical location 
across time and hence test models
of urban movement 
\cite{noulasurban}.

\bigskip

\noindent
\textbf{Sidebar 3}\\

\begin{center}
***********************************
\end{center}

Traag, Antonio and Van Dooren  \cite{socialbalance}
looked at the concept of \emph{social balance} (my friend's friend is my friend, my 
enemy's enemy is my friend, \ldots) to derive matrix-valued ordinary differential equations 
(ODEs) of the form 
$\dot A(t) = A(t) \times A(t)$ and 
$\dot A(t) = A(t) \times A(t)^T$.
Given such an $A(t)$, \cite{dyncen} developed an 
accompanying
ODE  for the level of importance, or \emph{centrality}, of the network nodes, 
showing that the matrix logarithm function arises naturally.
An alternative concept from the social sciences, 
\emph{triadic closure} (the more friends I have in common with somebody, 
the more likely I am to become their friend), was used in \cite{bistab} 
to derive a stochastic birth and death model for link dynamics.
There, a mean-field analysis agreed with simulations showing that the 
network can self-organize into either of two very different long-term behaviors.

\begin{center}
***********************************
\end{center}

\bigskip

In the preamble to his recent text \emph{The New Science of Cities} 
\cite{Batty2013}, Michael Batty,
from 
The Bartlett
Centre for Advanced Spatial Analysis, discusses three central principles
that inform his \lq\lq networks and flows\rq\rq\ perspective of city science; each of 
which resonates strongly with the standpoint of this article.
Batty's first principle is that the relations between objects, not the intrinsic 
attributes of those objects, should condition our understanding; a viewpoint 
familiar to those of us who have been exposed to graph theory or category theory. 
Second, we should aim to measure, categorize and look for universal scalings 
when we observe  and compare city networks across space and time.
Third, having gathered macrolevel observations we should seek 
to understand the micro-level principles that drive them---in the 
language of applied mathematics, we should aim for 
explanatory models, based on explicit modelling assumptions, 
with predictive power.
Batty's book 
makes use of concepts such as 
agent based modelling,
flocking,
graph theory, 
Markov chains,
Markovian decision problems,
optimization and 
self-similarity/fractals, and hence is an excellent starting-off point
for mathematicians wishing to enter the field.

In the spirit of micro-level digital interactions, one of us has 
 initiated a Linked-In group on 
\emph{MSSC: Mathematical Sciences for Smart Cities} 
 and interested readers are encouraged to join us.

\bigskip

\noindent
\textsf{Acknowledgement}
 
The authors are supported by 
the
Engineering and Physical Sciences 
Research Council Mathematical Sciences theme, 
the Research Council UK Digital Economy programme and 
the Royal Society/Wolfson Foundation.

\bibliographystyle{siam}
\bibliography{fcrefs}

\end{document}